\documentclass[preprint,amssymb,superscriptaddress]{revtex4}
\usepackage{amsmath}
\usepackage{amsthm}
\usepackage{amssymb}
\usepackage{amsfonts}
\usepackage{graphicx}
\usepackage{wasysym}
\usepackage{marvosym}
\usepackage{mathrsfs}

\newcommand{\ket}[1]{\left| #1 \right\rangle}

\newcommand{\avg}[1]{\text{Tr}\left\{ \widehat{\rho}~ #1 \right\}}

\newcommand{\opEps}{\widehat{E}^{(+)}}
\newcommand{\opEns}{\widehat{E}^{(-)}}
\newcommand{\Gpol}{\overleftrightarrow{G}^{(1)}(\mathbf{r},t;\mathbf{r},t)}
\newcommand{\Gpole}[1]{G_{#1}^{(1)}(\mathbf{r},t;\mathbf{r},t)}
\newcommand{\x}{\mathbf{r},t}

\newcommand{\pok}{\left|\psi_1\right\rangle}
\newcommand{\ptk}{\left|\psi_2\right\rangle}
\newcommand{\pob}{\left\langle \psi_1\right|}
\newcommand{\ptb}{\left\langle \psi_2\right|}
\newcommand{\I}{\mathscr{I}}
\newcommand{\kx}{\ket{1}_x\ket{0}_y}
\newcommand{\ky}{\ket{0}_x\ket{1}_y}

\newcommand{\oan}{\widehat{a}}

\begin{document}

\title{``Which-path information'' and partial polarization in single-photon interference experiments}

\author{Mayukh Lahiri}
\email{mayukh@pas.rochester.edu} \affiliation{Department of Physics
and Astronomy, University of Rochester, Rochester, NY 14627, U.S.A}


\begin{abstract}
\qquad It is shown that the degree of polarization of light,
generated by superposition in a single-photon interference
experiment, may depend on the indistinguishability of the
photon-paths. \vskip 1cm


\end{abstract}
\maketitle
\par
Quantum systems (quantons \cite{Note-quanton}) possess properties of
both particles and waves. Bohr's correspondence principle
\cite{Bohr} suggests that these two properties are mutually
exclusive. In other words, depending on the experimental situation,
a quanton will behave as a particle or as a wave. In the third
volume of his famous lecture series \cite{Feynman-lec-3}, Feynman
emphasized that this \emph{wave-particle duality} may be understood
from Young's two-pinhole type interference experiments
\cite{Young-interference}. In such an experiment, a quanton may
arrive at the detector along two different paths. If one can
determine which path the quanton traveled, then no interference
fringe will be found (i.e., the quanton will show complete particle
behavior). On the other hand, if one \emph{cannot} obtain any
information about the quanton's path, then interference fringes with
unit visibility will be obtained (i.e., the quanton will show
complete wave behavior), assuming that the intensities at the two
pinholes are the same. In the intermediate case when one has partial
``which-path information'' (WPI), fringes with visibility smaller
than unity are obtained, even if the intensities at the two pinholes
are equal. For the sake of brevity, we will use the term ``best
circumstances'' to refer to the situation when in an Young's
interference experiment, the intensities at the two pinholes are
equal, or to equivalent situations in other interferometric setups
\cite{Note-best-circums}. The relation between fringe visibility and
WPI has been investigated in Refs. \cite{M-ind,JSV,Eng}. It has been
established that a quantitative measure of WPI and fringe-visibility
obey a certain inequality \cite{Note-WPI-notation}.
\par
Aim of this paper is to show that it is not only the
fringe-visibility, but also the polarization properties of the
superposed light, which may depend on WPI in an interference
experiment. In particular, it will be established that the degree of
polarization and a measure of WPI obey an inequality, with equality
holding under the ``best circumstances''.
\par
Polarization properties of light have been extensively studied in
the framework of classical theory (see, for example,
\cite{BW,MW,B,Collett}). Foundation of this subject was laid down by
Stokes in 1852 in a classic paper \cite{GS} in which he formulated
the theory of polarization in terms of certain measurable
parameters, now known as Stokes parameters (see, for example,
\cite{MW}, p. 348). In the late 1920's, Wiener
\cite{NW,NW-1930,NWB}) showed that correlation matrices can be used
to analyze the polarization properties of light. Wolf \cite{W-pol}
used the correlation matrix formulation for systematic studies of
polarization properties of statistically stationary light beams
within the framework of classical theory. Later, some publications
addressed this problem by the use of quantum mechanical techniques
(see, for example, \cite{Fano,G-1,C-pol}). A discussion on quantum
mechanical analogues of the Stokes parameters can be found in Ref.
\cite{B}, Appendix L. Recently, another quantum mechanical analysis
of polarization properties of optical beams have been presented
\cite{LW-qp}.
\par
In this paper, we will mainly use the formulation of Ref.
\cite{LW-qp}. According to that formulation, the polarization
properties (based on first-order correlation functions
\cite{Note-first-second-order}) of light at a space-time point
($\x$) may be characterized by the so-called quantum-polarization
matrix
\begin{equation}\label{G-pol-elems}
\Gpol\equiv
\left[\Gpole{ij}\right]=\avg{\opEns_i(\x)\opEps_j(\x)},\quad
i=x,y;~~j=x,y,
\end{equation}
where $\opEps_i$ and $\opEns_i$ are the $i$ components of the
positive and of the negative frequency parts of the quantized
electric field operator respectively, and $\widehat{\rho}$
represents the density operator. A quantitative measure of
polarization properties of photons, at a space-time point ($\x$), is
given by the degree of polarization (\cite{LW-qp}, Eq. (31))
\begin{equation}\label{quant-deg-pol-def}
\mathscr{P}(\x)\equiv\sqrt{1-\frac{4~\text{Det}\Gpol}{\left\{\text{Tr}\Gpol\right\}^2}},
\end{equation}
where Det and Tr denote the determinant and the trace respectively.
This quantity is always bounded by zero and unity, i.e., $0\leq
\mathscr{P}(\x)\leq 1$. It is to be noted that $\mathscr{P}(\x)$ is
expressed in terms of the trace and the determinant of the matrix
$\Gpol$, and hence is invariant under unitary transformations. When
$\mathscr{P}(\x)=0$, the light is completely unpolarized at the
space-time point $(\x)$ and when $\mathscr{P}(\x)=1$, it is
completely polarized at that space-time point. In intermediate cases
($0<\mathscr{P}(\x)<1$), the light is said to be partially
polarized.
\par
Suppose, now, that $\pok$ and $\ptk$ represent two normalized
single-photon states (eigenstates of the number operator), so that
\begin{subequations}\label{state-props}
\begin{align}
\left\langle \psi_1 | \psi_2\right\rangle &=\left\langle \psi_2 | \psi_1 \right\rangle=0, \label{state-props:a}\\
\left\langle \psi_1 | \psi_1\right\rangle &=\left\langle \psi_2 |
\psi_2 \right\rangle=1 \label{state-props:b}.
\end{align}
\end{subequations}
We first consider a state $\ket{\psi_{\text{ID}}}$ of light, which
is formed by coherent superposition of the two states $\pok$ and
$\ptk$, i.e.,
\begin{equation}\label{st-coh-sup}
\ket{\psi_{\text{ID}}}=\alpha_1\pok+ \alpha_2\pok, \quad
|\alpha_1|^2+|\alpha_2|^2=1,
\end{equation}
where $\alpha_1$ and $\alpha_2$ are, in general, two complex
numbers. In this case, a photon may be in the state $\pok$ with
probability $|\alpha_1|^2$, or in the state $\ptk$ with probability
$|\alpha_2|^2$, but the two possibilities are intrinsically
\emph{indistinguishable}. The density operator
$\widehat{\rho}_{\text{ID}}$ will then have the form
\begin{equation}\label{den-op-ID}
\widehat{\rho}_{\text{ID}}=|\alpha_1|^2 \pok\pob+|\alpha_2|^2
\ptk\ptb+ \alpha_1^{\ast}\alpha_2 \ptk\pob+ \alpha_2^{\ast}\alpha_1
\pok\ptb.
\end{equation}
In the other extreme case, when the state of light is due to
incoherent superposition of the two states, the density operator
$\widehat{\rho}_{\text{D}}$ will be given by the expression
\begin{equation}\label{den-op-D}
\widehat{\rho}_{\text{D}}=|\alpha_1|^2 \pok\pob+|\alpha_2|^2
\ptk\ptb.
\end{equation}
Here $|\alpha_1|^2$ and $|\alpha_2|^2$ again represent the
probabilities that the photon will be in state $\pok$ or in state
$\ptk$, but now the two possibilities are intrinsically
\emph{distinguishable}. Mandel \cite{M-ind} showed that in any
intermediate case, the density operator
\begin{equation}\label{den-op-gen}
\widehat{\rho}=\rho_{11} \pok\pob+ \rho_{12}\pok\ptb+ \rho_{21}
\ptk\pob+\rho_{22} \ptk\ptb
\end{equation}
can be uniquely expressed in the form
\begin{equation}\label{den-op-decom}
\widehat{\rho}=\I\widehat{\rho}_{\text{ID}}+(1-\I)\widehat{\rho}_{\text{D}},
\qquad \qquad 0\leq \I \leq 1.
\end{equation}
Mandel's defined $\I$ as the \emph{degree of indistinguishability}.
If $\I=0$, the two paths are completely distinguishable, i.e., one
has complete WPI; and if $\I=1$, they are completely
indistinguishable, i.e., one has no WPI. In the intermediate case
$0<\I<1$, the two possibilities may be said to be partially
distinguishable. Clearly, $\I$ may be considered as a measure of
WPI. According to Eqs. (\ref{den-op-gen}) and (\ref{den-op-decom}),
one can always express $\widehat{\rho}$ in the form
\begin{equation}\label{den-op-gen-form}
\widehat{\rho}=|\alpha_1|^2 \pok\pob+|\alpha_2|^2 \ptk\ptb+
\I(\alpha_1^{\ast}\alpha_2\ptk\pob+ \alpha_2^{\ast}\alpha_1
\pok\ptb).
\end{equation}
Clearly, the condition of ``best circumstances'' requires that
$|\alpha_1|=|\alpha_2|$.
\par
For the sake of simplicity, let us assume that $\pok$ and $\ptk$ are
of the form
\begin{subequations}\label{psi-form}
\begin{align}
\pok &=\kx, \label{psi-form:a} \\
\ptk &=\ky, \label{psi-form:a}
\end{align}
\end{subequations}
where the two states are labeled by the same (vector) mode
$\mathbf{k}$, and $x$, $y$ are two mutually orthogonal directions,
both perpendicular to the direction of $\mathbf{k}$ [for the sake of
brevity, $\mathbf{k}$ is not displayed in Eqs. (\ref{psi-form})].
Clearly $\pok$ represents the state of a photon polarized along the
$x$ direction, and $\ptk$ represents that along the $y$ direction.
In the present case, one may express $\opEps_i(\x)$ in the form
\begin{equation}\label{E-p-form}
\opEps_i(\x)=Ce^{i(\mathbf{k}\cdot \mathbf{r}-\omega
t)}\oan_i,\qquad (i=x,y),
\end{equation}
where the operator $\oan_i$ represents annihilation of a photon in
mode $\mathbf{k}$, polarized along the $i-$ axis, and $C$ is a
constant. From Eqs. (\ref{G-pol-elems}), (\ref{den-op-gen-form}),
and (\ref{E-p-form}), one readily finds that the quantum
polarization matrix $\Gpol$ has the form
\begin{equation}\label{G-pol-form}
\Gpol=|C|^2\begin{pmatrix}
|\alpha_1|^2 & \I\alpha_1^{\ast}\alpha_2 \\
\I\alpha_1\alpha_2^{\ast} & |\alpha_2|^2 \hfill
\end{pmatrix}.
\end{equation}
From Eqs. (\ref{quant-deg-pol-def}) and (\ref{G-pol-form}) and using
the fact $|\alpha_1|^2+|\alpha_2|^2=1$, one finds that, in this
case, the degree of polarization is given by the expression
\begin{equation}\label{deg-pol-calc}
\mathscr{P}=\sqrt{(|\alpha_1|^2-|\alpha_2|^2)^2+4|\alpha_1|^2|\alpha_2|^2\I^2}.
\end{equation}
\par
It follows from Eq. (\ref{deg-pol-calc}) by simple calculations that
\begin{equation}\label{P-I-ineq-int-step}
\mathscr{P}^2-\I^2=(1-\I^2)(2|\alpha_1|^2-1)^2.
\end{equation}
Using the fact that $0\leq \I \leq 1$, one readily finds that
\begin{equation}\label{P-I-ineq}
\mathscr{P}\geq\I.
\end{equation}
Thus, the degree of polarization of the out-put light in a
single-photon interference experiment is always greater or equal to
the degree of indistinguishability ($\I$) which a measure of
``which-path information''.
\par
Let us now assume that the condition of ``best circumstances'' is
achieved, i.e., one has $|\alpha_1|^2=|\alpha_2|^2$. It then readily
follows from Eq. (\ref{deg-pol-calc}) that
\begin{equation}\label{deg-pol-eq-indist}
\mathscr{P}=\I.
\end{equation}
This formula shows that under the ``best circumstances'', the degree
of indistinguishability (a measure of WPI) and the degree of
polarization are equal. We will now discuss the physical
interpretation of this result. If one has complete ``which-path
information'' (i.e., $\I=0$), then from Eq.
(\ref{deg-pol-eq-indist}) it follows that the degree of polarization
of the light emerging from the interferometer is equal to zero.
Complete ``which-path information'' in an single-photon interference
experiment implies that a photon shows complete particle nature, and
our analysis suggests that in such a case light is completely
unpolarized. In the other extreme case, when one has no ``which-path
information'', i.e., when a photon does \emph{not} display any
particle behavior, the output light will be completely polarized.
Any intermediate case will produce partially polarized light. It is
clear that such a phenomenon is completely quantum-mechanical and
cannot be realized by the use of classical theory.
\par
We summarize our result by saying that in a single-photon experiment
the polarization properties of the light emerging from the
interferometer depend on whether a photon behaves like a particle or
like a wave.
\par
I am indebted to Professor Emil Wolf for critical reading of this
manuscript and helpful suggestions. I thank Mr. Shantanu Agarwal for
drawing my attention to the important papers cited as Refs.
\cite{JSV,Eng}. The research was supported by the US Air Force
Office of Scientific Research under grant No. FA9550-08-1-0417.


\newpage
\begin{thebibliography}{99}
\section*{References and Notes}

\bibitem {Note-quanton} This abbreviation is due to M. Bunge [see,
for example, J.-M. L\'evy-Leblond, Physica B \textbf{151}, 314
(1988), and also B.-G. Englert, Phys. Rev. Lett. \textbf{77}, 2154
(1996)].

\bibitem {Bohr} N. Bohr, Naturwissenschaften \textbf{16}, 245
(1928).

\bibitem {Feynman-lec-3} R. P. Feynman, R. B. Leighton and M. Sands, \textit{The Feynman
Lectures on Physics}, Vol. III, (New York, Addison-Wesley Publishing
Company, 1966).

\bibitem {Young-interference} T. Young, ``An account of some
cases of the production of colours, not hitherto described'', Phil.
Trans. Roy. Soc. Lond., \textbf{92}, 387-397 (1802).

\bibitem {Note-best-circums} The term ``best circumstances'' was
used by Zernike in a classic paper \cite{Z} on coherence theory. By
this term, he meant that the intensities of the two interfering
beams were equal and that only small path difference was introduced
between them. In our analysis, we do not consider any path
difference; thus in our case the term ``best circumstances'' implies
that the intensities at the two pinholes, in an Young's interference
experiment, are equal to each other (or equivalent situations in
other interference experiments).

\bibitem {Z} F. Zernike, ``The concept of degree of cohernence and its
application and its application to optical problems'',
\textit{Physica} \textbf{5}, 785 (1938).

\bibitem {M-ind} L. Mandel, ``Coherence and
indistinguishability'', Opt. Lett. \textbf{16}, 1882-1883 (1991).

\bibitem {JSV} G. Jaeger, A. Shimony and L. Vaidman, ``Two
interferometric complementarities'', Phys. Rev. A \textbf{51}, 54–67
(1995).

\bibitem {Eng} B-G. Englert, ``Fringe vsibility and which-way
information: an inequality'', Phys. Rev. Lett. \textbf{77},
2154-2157 (1996).

\bibitem {Note-WPI-notation} WPI has been represented by different
quantities in \cite{M-ind} and in \cite{Eng}. Their relationship is
discussed in Note [5] of Ref. \cite{Eng}.

\bibitem {BW} M. Born and E. Wolf, \textit{Principles of
Optics} (Cambridge University Press, Cambridge, 7th Ed. 1999).

\bibitem {MW} L. Mandel and E. Wolf, \textit{Optical Coherence and
Quantum Optics} (Cambridge, Cambridge University Press, 1995).

\bibitem {B} C. Brosseau, \textit{Fundamentals of Polarized Light: A Statistical
Optics Approach} (Wiley, 1995).

\bibitem {Collett} E. Collett \textit{Polarized light : fundamentals and
applications} (New York, Marcel Dekker, 1993).

\bibitem {GS} G. G. Stokes, \textit{Trans. Cambr. Phil. Soc.} \textbf{9}, 399 (1852).

\bibitem {NW} N. Wiener, ``Coherency matrices and quantum theory'', J. Math. Phys. (M.I.T.) \textbf{7}, 109-125 (1928).

\bibitem {NW-1930} N. Wiener, ``Generalized harmonic analysis'', Acta Math. \textbf{55},
117-258 (1930).

\bibitem {NWB} N. Wiener, \textit{Generalized harmonic analysis and Tauberian
theorems} (The M.I.T. Press, First paperback edition, 1966).

\bibitem {W-pol} E. Wolf, ``Coherence Properties of Partially Polarized Electromagnetic Radiation'', Nuovo
Cimento \textbf{13}, 1165–1181 (1959).

\bibitem {Fano} U. Fano, ``A Stokes-parameter technique for the
treatment of polarization in quantum mechanics'', Phys. Rev.,
\textbf{93}, 121-123 (1954).

\bibitem {G-1} R. J. Glauber, ``The quantum theory of optical coherence''
Phys. Rev., \textbf{130}, 2529-2539 (1963).

\bibitem {C-pol} E. Collett, Am. J. Phys. \textbf{38}, 563 (1970).

\bibitem {LW-qp} M. Lahiri and E. Wolf, ``Quantum analysis of
polarization properties of optical beams,'' Phys. Rev. A
\textbf{82}, 043805 (2010).

\bibitem {Note-first-second-order} Analogous correlation
functions in the classical theory are referred as second-order ones.

\end{thebibliography}
\end{document}